\RequirePackage{lineno}
\documentclass[prd,amsmath,amssymb,showpacs,superscriptaddress,nofootinbib,twocolumn]{revtex4-1}
\usepackage{graphicx}
\usepackage{epsfig}
\usepackage{dcolumn}
\usepackage{bm}
\usepackage{overpic}
\usepackage{longtable}
\usepackage{lineno}
\usepackage{color}  
\usepackage{multirow}
\usepackage[colorlinks,linkcolor=blue,anchorcolor=blue,citecolor=blue]{hyperref} 

\begin{document}
\normalsize

\parskip=5pt plus 1pt minus 1pt

\title{\boldmath Luminosity measurements for the $R$ scan experiment at BESIII}

\author{
  \begin{small}
    \begin{center}
M.~Ablikim$^{1}$, M.~N.~Achasov$^{9,e}$, S. ~Ahmed$^{14}$, X.~C.~Ai$^{1}$, O.~Albayrak$^{5}$, M.~Albrecht$^{4}$, D.~J.~Ambrose$^{44}$, A.~Amoroso$^{49A,49C}$, F.~F.~An$^{1}$, Q.~An$^{46,a}$
% ...
      \\
      \vspace{0.2cm}
      (BESIII Collaboration)\\
      \vspace{0.2cm} {\it
$^{1}$ Institute of High Energy Physics, Beijing 100049, People's Republic of China\\
% ...
$^{4}$ Bochum Ruhr-University, D-44780 Bochum, Germany\\
$^{5}$ Carnegie Mellon University, Pittsburgh, Pennsylvania 15213, USA\\
% ...
$^{9}$ G.I. Budker Institute of Nuclear Physics SB RAS (BINP), Novosibirsk 630090, Russia\\
% ...
$^{14}$ Helmholtz Institute Mainz, Johann-Joachim-Becher-Weg 45, D-55099 Mainz, Germany\\
% ...
$^{44}$ University of Rochester, Rochester, New York 14627, USA\\
% ...
$^{46}$ University of Science and Technology of China, Hefei 230026, People's Republic of China\\
% ...
$^{49}$ (A)University of Turin, I-10125, Turin, Italy; (B)University of Eastern Piedmont, I-15121, Alessandria, Italy; (C)INFN, I-10125, Turin, Italy\\
% ...
$^{a}$ Also at State Key Laboratory of Particle Detection and Electronics, Beijing 100049, Hefei 230026, People's Republic of China\\
% ...
$^{e}$ Also at the Novosibirsk State University, Novosibirsk, 630090, Russia\\
% ...
      }\end{center}
    \vspace{0.4cm}
  \end{small}
}

\affiliation{}


\begin{abstract}

\textbf{Abstract}: By analyzing the large-angle Bhabha scattering events $e^{+}e^{-}$ $\to$ ($\gamma$)$e^{+}e^{-}$ and diphoton events $e^{+}e^{-}$ $\to$ $\gamma\gamma$ for the data sets collected at center-of-mass (c.m.) energies between 2.2324 and 4.5900 GeV (131 energy points in total) with the upgraded Beijing Spectrometer (BESIII) at the Beijing Electron-Positron Collider (BEPCII), the integrated luminosities have been measured at the different c.m. energies, individually. The results are the important inputs for R value and $J/\psi$ resonance parameter measurements.


\end{abstract}

\maketitle

\section{\boldmath INTRODUCTION}
Hadron production in $e^{+}e^{-}$ annihilation is one of the most valuable testing grounds for Quantum Chromodynamics (QCD), and is an important input for precision tests of the Standard Model (SM). The R value, which is defined as the lowest level hadronic cross section normalized by the theoretical $\mu^{+}\mu^{-}$ production cross section in $e^{+}e^{-}$ annihilation, is an indispensable input for the determination of the non-perturbative hadronic contribution to the electromagnetic coupling constant evaluated at the Z pole ($\alpha(M_{Z}^{2})$)~\cite{R_cal1,R_cal2}, and the anomalous magnetic moment $a_{\mu}$ $=$ $(g-2)/2$ of the muon~\cite{R_cal3}. The dominant uncertainties in both $\alpha(M_{Z}^{2})$ and $a_{\mu}$ measurements are due to the effects of hadronic vacuum polarization, which cannot be reliably calculated in the low energy region. Instead, with the application of dispersion relations, experimentally measured R values can determine the effect of vacuum polarization.

In experiment, the R value is determined by
\begin{equation}
  {\rm R} = \frac{N_{\rm had}^{\rm obs}-N_{\rm had}^{\rm bkg}}{\sigma_{\mu\mu}^{0}\cdot\mathcal{L}\cdot\varepsilon_{\rm had}\cdot\varepsilon_{\rm had}^{\rm trig}\cdot(1+\delta)},
\end{equation}
where $N_{\rm had}^{\rm obs}$ is the number of observed hadronic events, $N_{\rm had}^{\rm bkg}$ is the number of background events, $\mathcal{L}$ is the integrated luminosity, $\varepsilon_{\rm had}$ is the detection efficiency for the hadron event selection, $\varepsilon_{\rm had}^{\rm trig}$ is the trigger efficiency, $1+\delta$ is the initial state radiation (ISR) correction factor, and $\sigma_{\mu\mu}^{0}$ is the Born cross section of $e^{+}e^{-} \to \mu^{+}\mu^{-}$. Therefore, the measurement of integrated luminosity plays an important role in the R value measurement.

Quantum electrodynamics (QED) processes can usually be used to determine the integrated luminosity due to larger production rates, simpler final state topologies and more accurate cross section calculation in theory relative to the other processes. The integrated luminosity is measured by
\begin{equation}
    \mathcal{L}=\frac{N_{\rm QED}^{\rm obs}-N_{\rm QED}^{\rm bkg}}{\sigma_{\rm QED} \cdot \varepsilon_{\rm QED} \cdot \varepsilon_{\rm QED}^{\rm trig}},
\end{equation}
where $N_{\rm QED}^{\rm obs}$ is the number of the QED events observed in the experimental data, $N_{\rm QED}^{\rm bkg}$ is the number of background events, $\sigma_{\rm QED}$ is the cross section of the selected QED process, $\varepsilon_{\rm QED}$ is the detection efficiency and $\varepsilon_{\rm QED}^{\rm trig}$ is the trigger efficiency.

In this paper, we present the measurements of lumonisities of the R scan data samples taken at BESIII from 2012 to 2014. The measurements are performed by analyzing two QED processes $e^{+}e^{-}\to$ ($\gamma$)$e^{+}e^{-}$ and $e^{+}e^{-}\to \gamma\gamma$. For energy points near the $J/\psi$ resonance, only the $e^{+}e^{-}$ $\to$ $\gamma\gamma$ process is used, because $J/\psi$ $\to$ ($\gamma$)$e^{+}e^{-}$ events can not be distinguished from $e^{+}e^{-}$ $\to$ ($\gamma$)$e^{+}e^{-}$ events experimentally.

\section{\boldmath DETECTOR}
BEPCII~\cite{NIM:DET} is a double-ring $e^{+}e^{-}$ collider designed to provide a peak luminosity of $10^{33}$ $\rm cm^{-2}\rm s^{-1}$ at the center-of-mass (c.m.) energy ($\sqrt{s}$) of $3770$ MeV. The BESIII~\cite{NIM:DET} detector has a geometrical acceptance of $93\%$ of $4\pi$ and has four main detector sub-components: (1) A small-cell, helium-based ($60\%$ He, $40\%$ C$_{3}$H$_{8}$) main drift chamber (MDC) with $43$ layers providing an average single-hit resolution of $135$ $\mu$m, and charged-particle momentum resolution in a $1$ T magnetic field of $0.5\%$ at $1$ GeV$/c$. (2) An electromagnetic calorimeter (EMC) consisting of 6240 CsI(Tl) crystals in cylindrical structure arranged in a barrel and two end-caps. The energy resolution at $1.0$ GeV$/c$ is $2.5\%$ ($5\%$) in the barrel (endcaps), and the position resolution is $6$ mm ($9$ mm) in the barrel (endcaps). (3) A time-of-flight (TOF) system for particle identification composed of a barrel part made of two layers with 88 pieces of 5~cm thick, 2.4~m long plastic scintillators in each layer, and two endcaps with 96 fan-shaped, 5 cm thick, plastic scintillators in each endcap. The time resolution of $80$ ps ($110$ ps) for barrel (endcap) prodvides $2\sigma$ $K/\pi$ separation for momenta up to $\sim 1.0$ GeV$/c$.  (4) A muon system (MUC) consisted of $1000$ m$^{2}$ of resistive plate chambers in nine (eight) layers of barrel (endcap) provides $2$ cm position resolution.

\section{\boldmath DATA SAMPLE AND MONTE CARLO SIMULATION}
The measurements of luminosities are performed for 131 data samples, including 4 energy points at 2.2324, 2.4000, 2.8000, 3.4000 GeV taken at the 2012 run, 104 energy points from 3.8500 to 4.5900 GeV taken at the 2013--2014 runs, 15 energy points near the $J/\psi$ production threshold, 4 energy points during the $\tau$ mass measurement and 4 energy points for charmonium studies.

The $e^{+}$$e^{-}$ $\to$ ($\gamma$)$e^{+}$$e^{-}$, $\gamma$$\gamma$ and ($\gamma$)$\mu^{+}$$\mu^{-}$ events are simulated with the generator Babayaga v$3.5$~\cite{babayaga}. The background process of $e^{+}$$e^{-}$ $\to$ $\tau^{+}$$\tau^{-}$ is generated with the KKMC~\cite{KKMC}, while the $e^{+}$$e^{-}$ $\to$ hadrons and $e^{+}$$e^{-}$ $\to$ $e^{+}$$e^{-}$ + X (X can be hadrons or leptons) events are generated with LUARLW~\cite{LUND2} and BesTwogam~\cite{bestwogam}, respectively.

\section{\boldmath ANALYSIS}
\begin{figure*}[htbp]
\begin{center}
\begin{overpic}[width=6.3cm,height=5cm,angle=0]{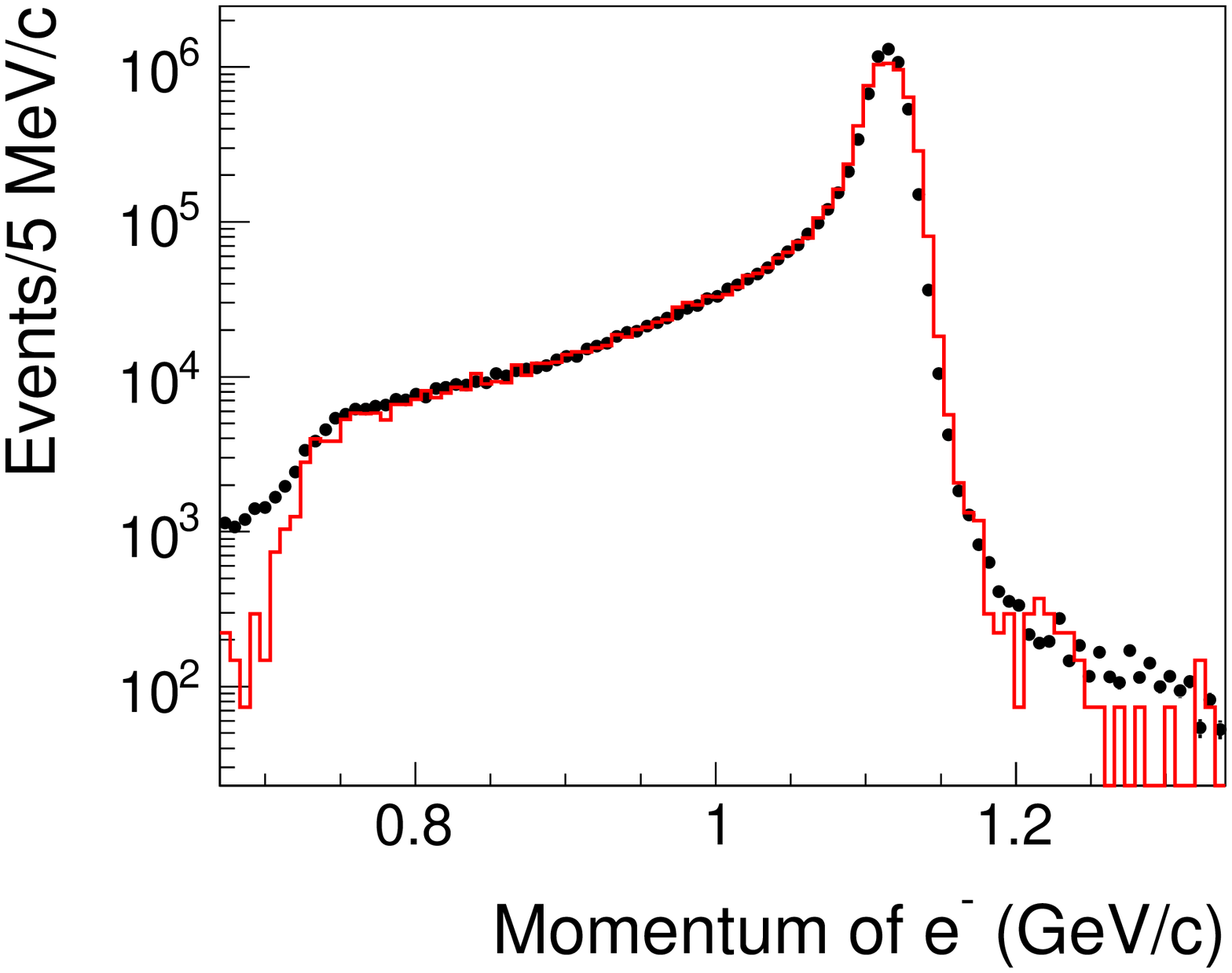}
\put(50,30){}
\end{overpic}
\begin{overpic}[width=6.3cm,height=5cm,angle=0]{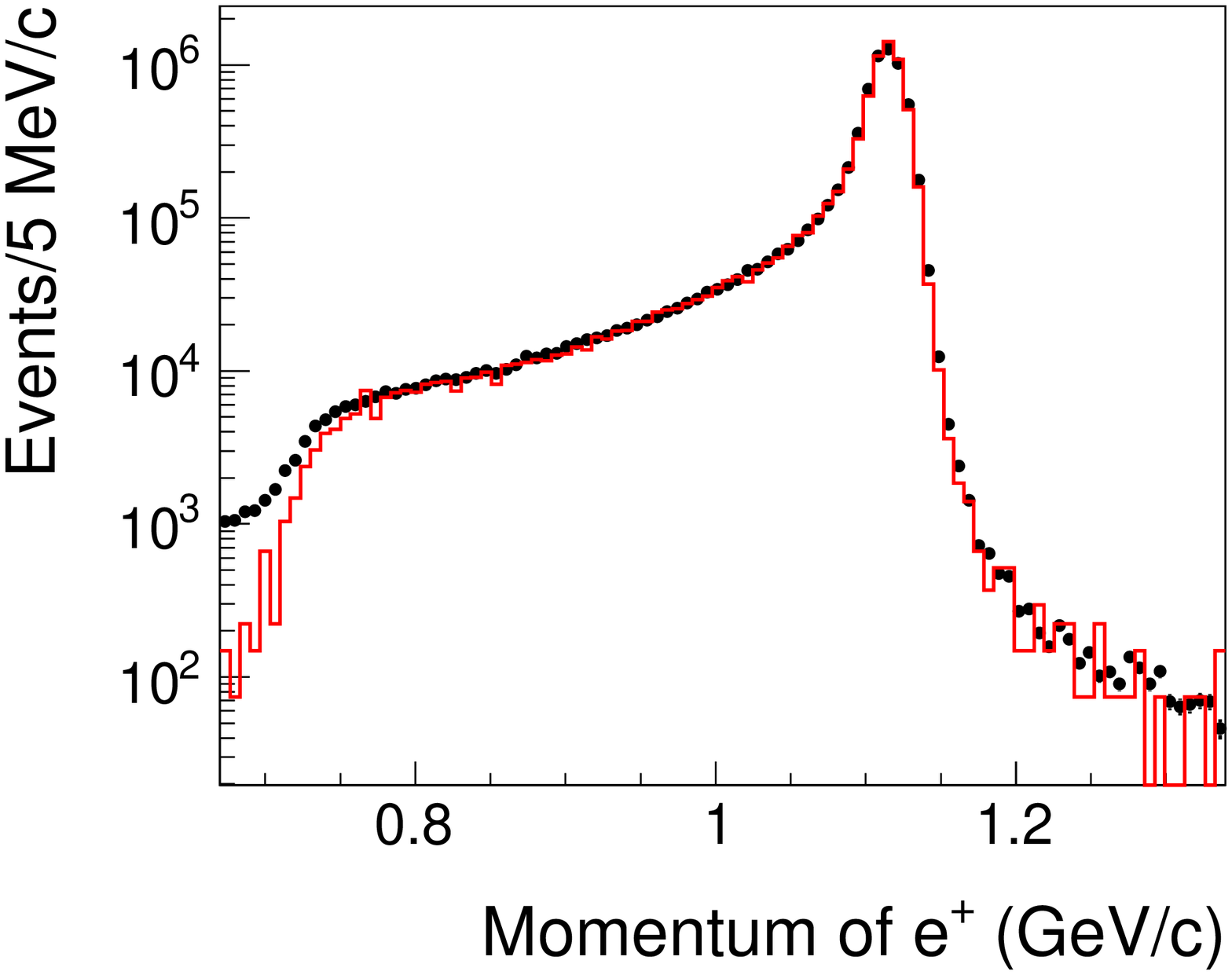}
\put(50,30){}
\end{overpic}
\begin{overpic}[width=6.3cm,height=5cm,angle=0]{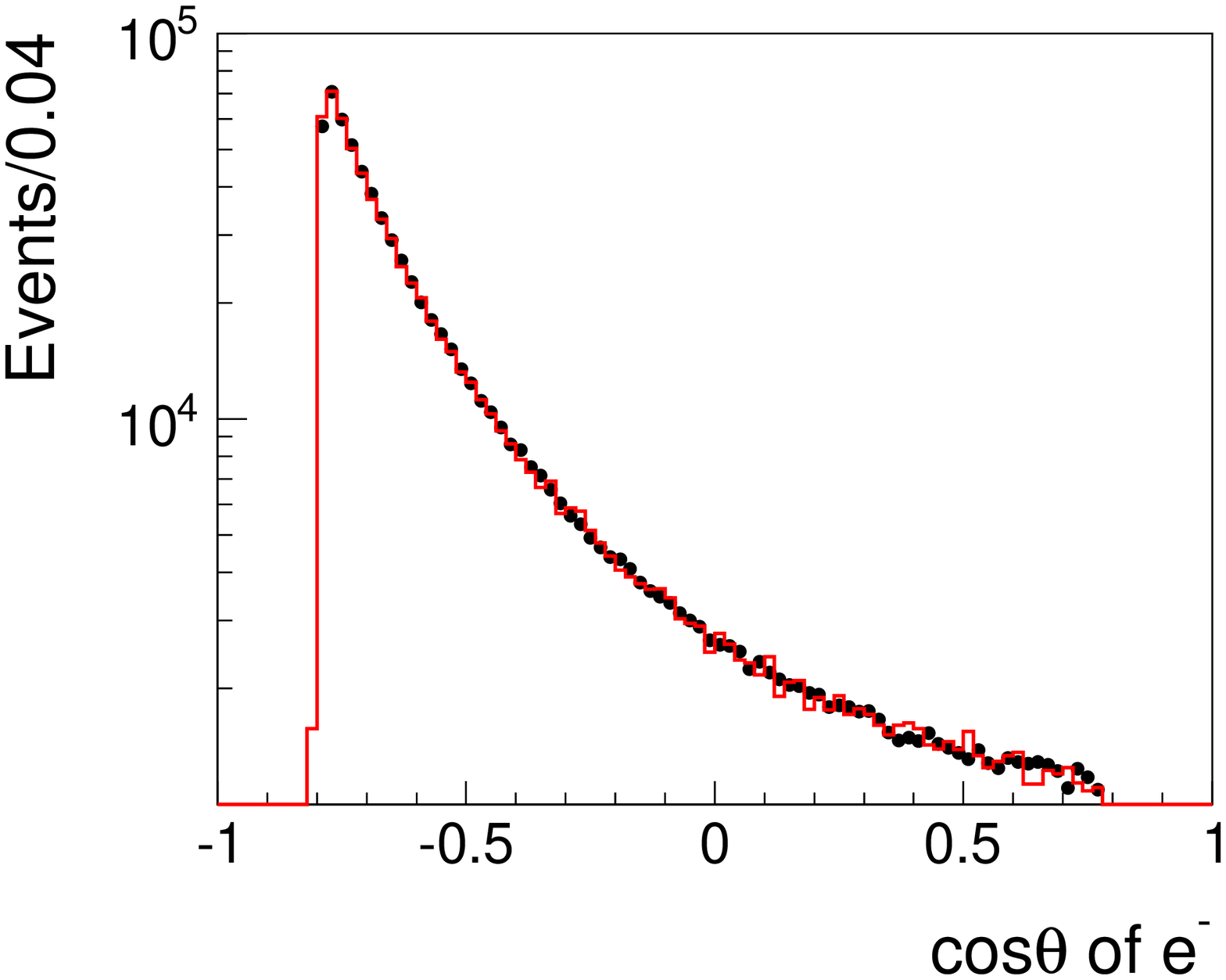}
\put(50,40){}
\end{overpic}
\begin{overpic}[width=6.3cm,height=5cm,angle=0]{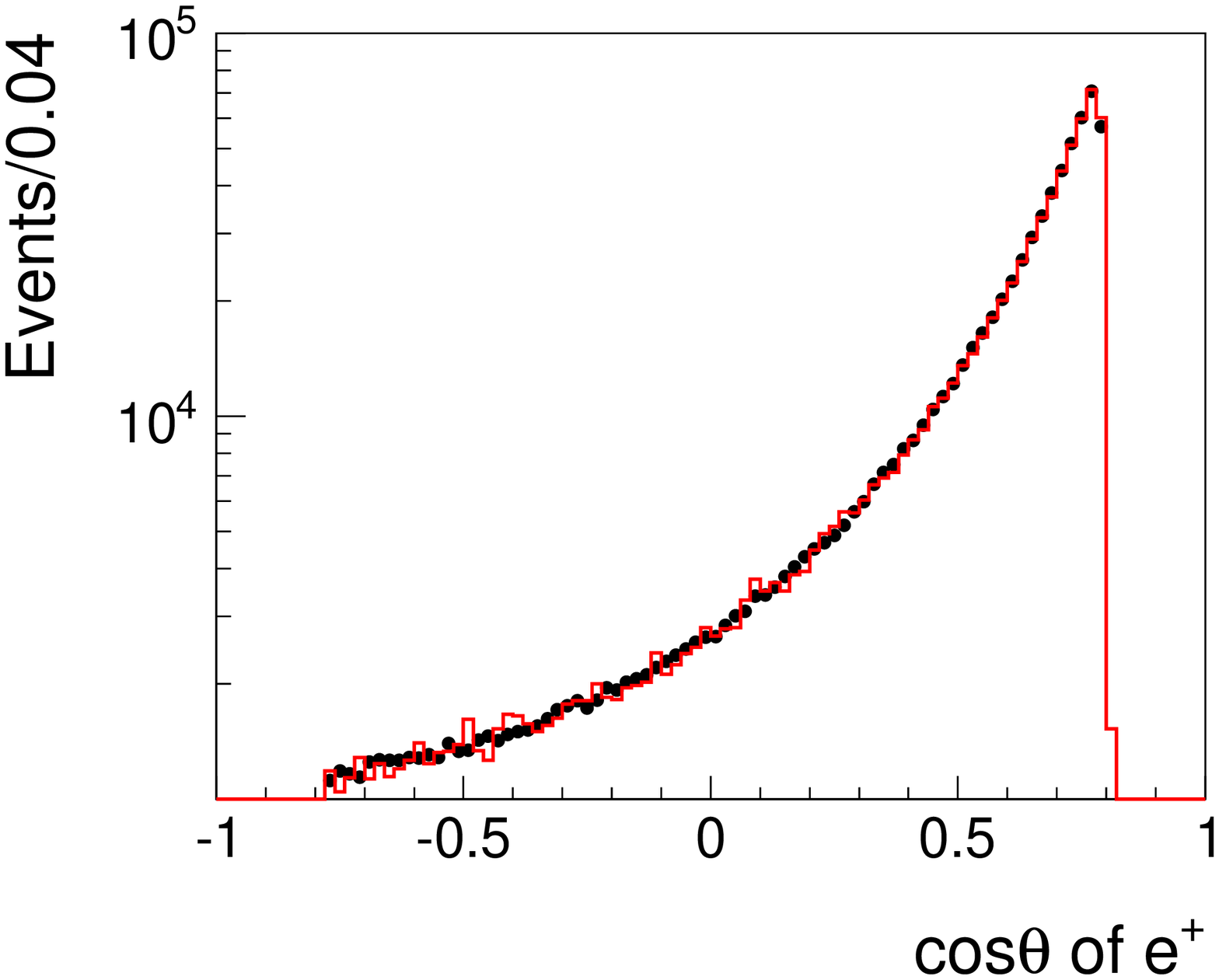}
\put(50,40){}
\end{overpic}
\end{center}
\caption{ The distributions of momentum (up plots) and polar angle $\cos\theta$ (down plots) for electron (left) and positron (right) at $\sqrt{s}$ = 2.2324 GeV. Dots with error bars are experimental data and red histograms are signal MC simulation. The MC entries are normalized to the experimental data.}\label{compare_bhabha}
\end{figure*}

The $e^{+}$$e^{-}$ $\to$ ($\gamma$)$e^{+}$$e^{-}$ events are required to have two good charged tracks with opposite charge. Each charged track is required to be within $\pm$10 cm of the interaction point in the beam direction and 1 cm in the plane perpendicular to the beam. In addition, the charged tracks are required to be within $|\cos\theta|$ $<$ 0.8, where $\theta$ is the polar angle, in the MDC. Without applying further particle identification, the tracks are assigned as electron and positron depending on their charges. The deposited energies of electron and positron ($E_{e^{\pm}}$) in the EMC are required to be larger than $0.65$ $\times$ $E_{\rm beam}$ to suppress backgrounds, where $E_{\rm beam}$ is the beam energy. To make sure the the selected charged tracks are back-to-back in the c.m. system, $|\Delta\theta_{e^{\pm}}|$ $=$ $|\theta_{1}+\theta_{2}-180^{\circ}|$ $<$ $10.0^{\circ}$ and $|\Delta\phi_{e^{\pm}}|$ $=$ $||\phi_{1}-\phi_{2}|-180^{\circ}|$ $<$ $5.0^{\circ}$ are required, where $\theta_{1/2}$ and $\phi_{1/2}$ are the polar and azimuthal angles of the two charged tracks, respectively. Figure~\ref{compare_bhabha} shows the comparisons of the momentum and polar angle distributions of electron and positron between experimental data and Monte Carlo (MC) simulation at $\sqrt{s}$ = 2.2324 GeV, the good agreements are observed.

\begin{figure*}[htbp]
\begin{center}
\begin{overpic}[width=6.3cm,height=5cm,angle=0]{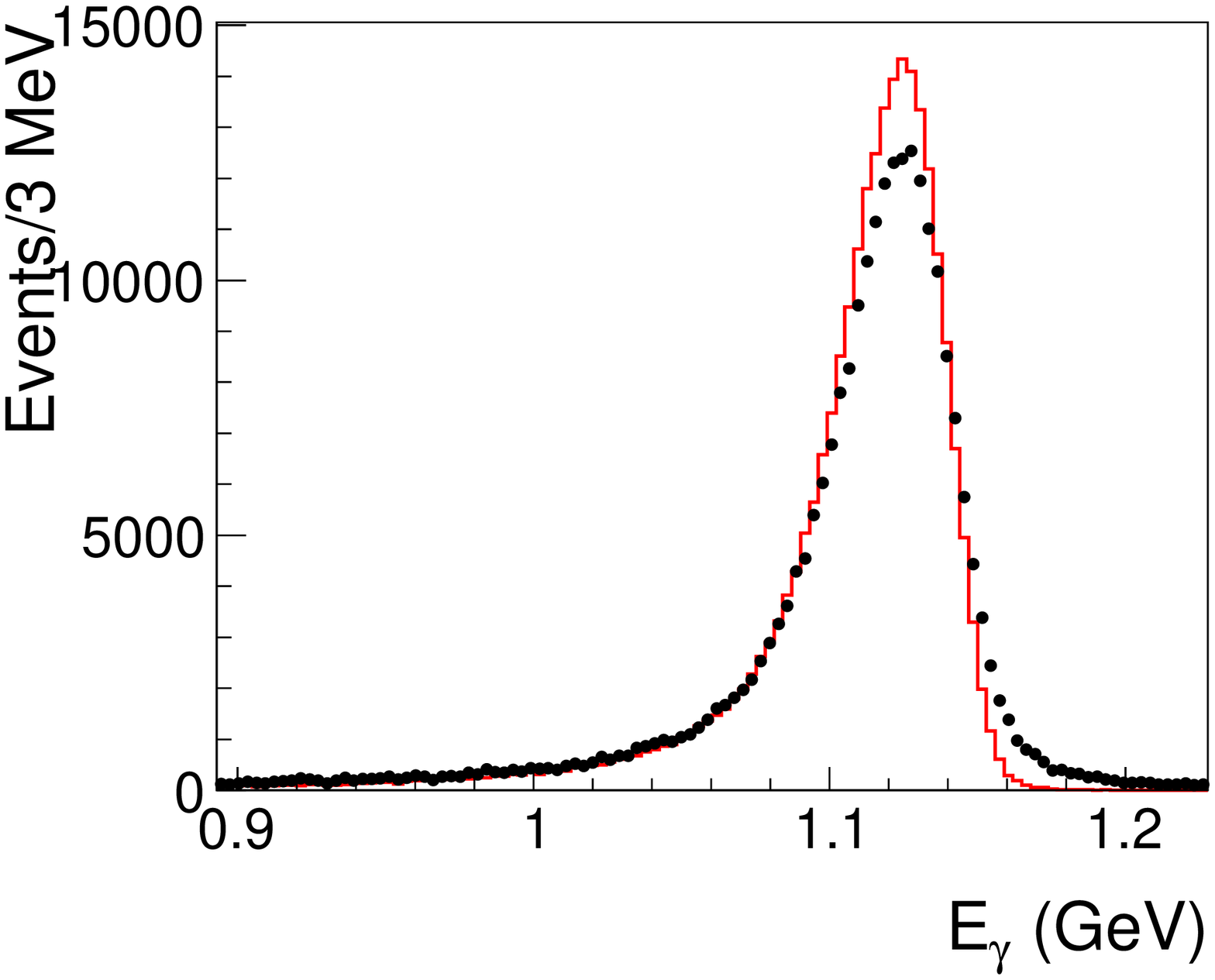}
\put(60,60){}
\end{overpic}
\begin{overpic}[width=6.3cm,height=5cm,angle=0]{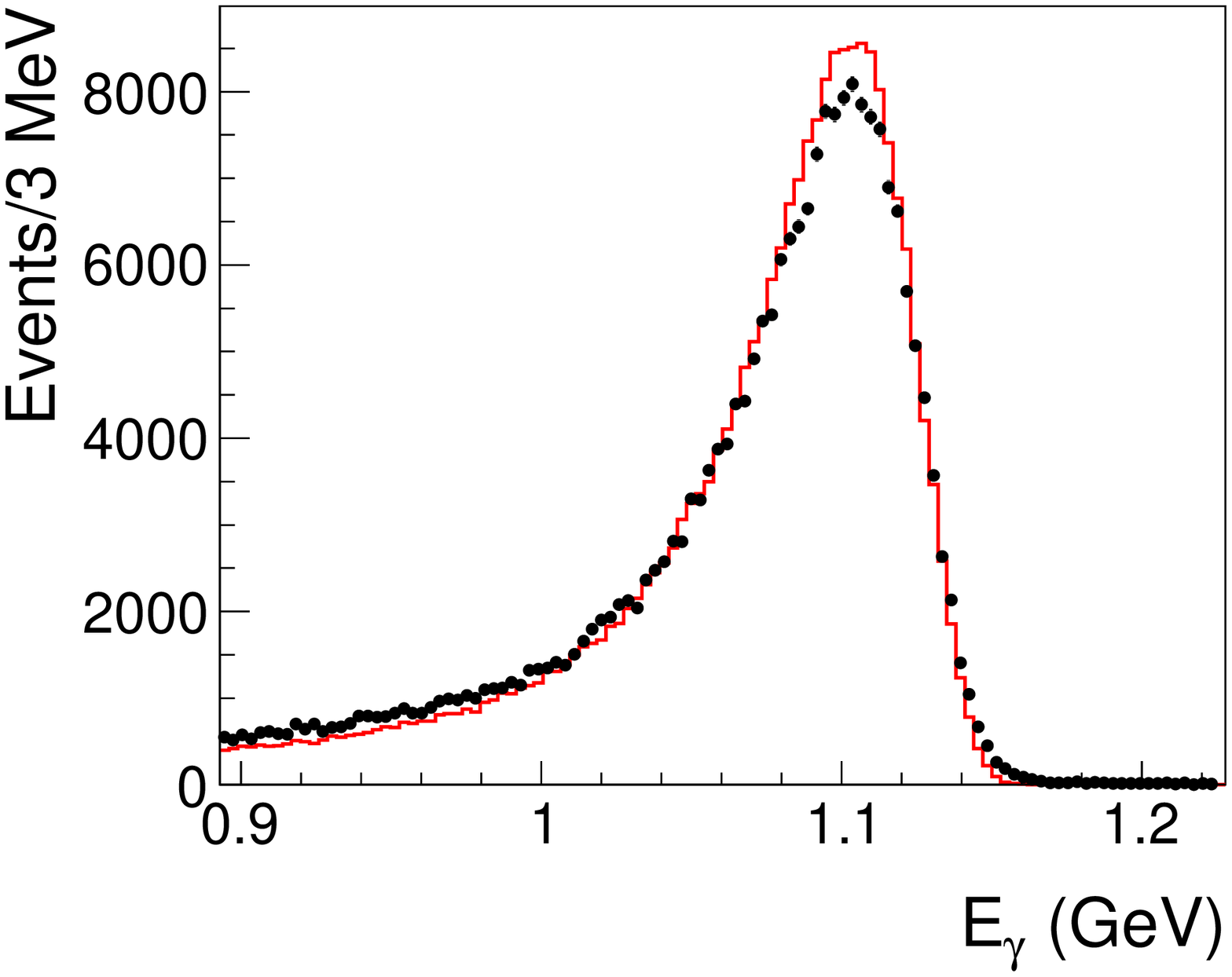}
\put(60,60){}
\end{overpic}
\begin{overpic}[width=6.3cm,height=5cm,angle=0]{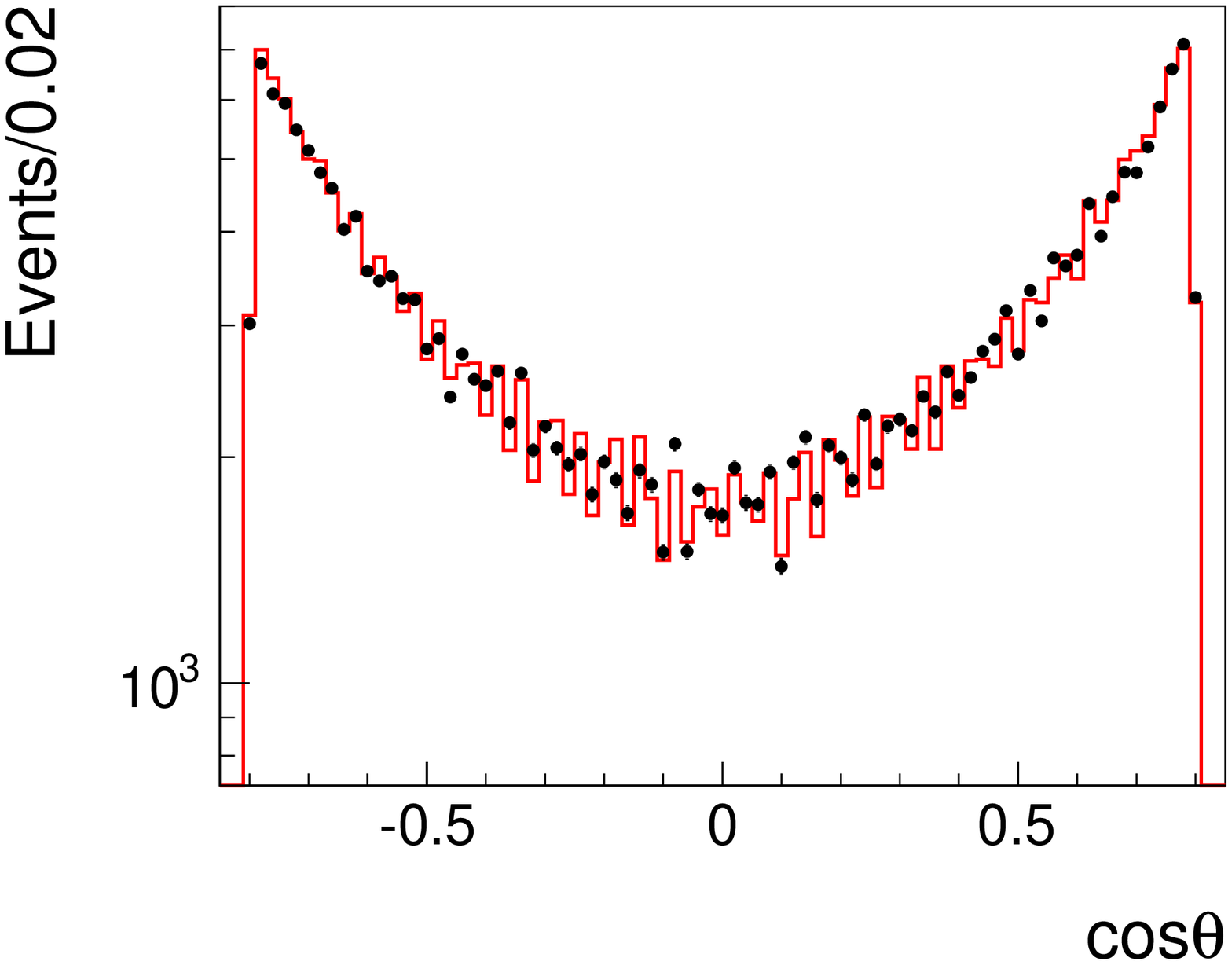}
\put(65,60){}
\end{overpic}
\begin{overpic}[width=6.3cm,height=5cm,angle=0]{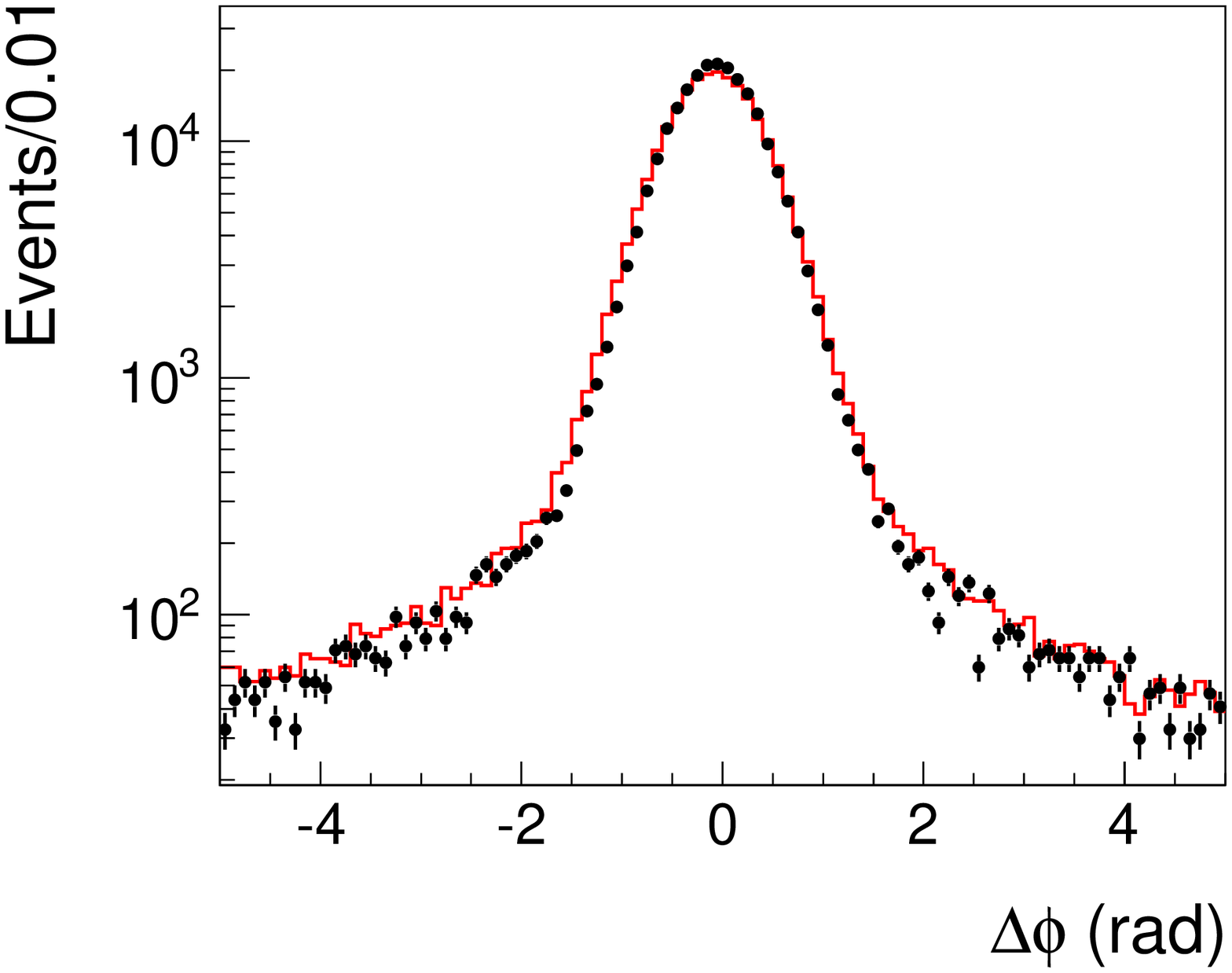}
\put(70,60){}
\end{overpic}
\end{center}
\caption{ Deposited energy distributions of the most energetic $\gamma$ (upper left), the secondary most energetic $\gamma$ (upper right), $\rm cos\theta$ (bottom left) and $\Delta\phi$ (bottem right) at $\sqrt{s}$ = 2.2324 GeV. Dots with error bars are experimental data and red histograms are signal MC simulation. The MC entries are normalized to the experimental data. The discrepancy in the deposited energy distributions is due to the imperfect simulation of energy correction deposited in TOF. However, it will not affect the efficiency since loose requirements on these variables are applied. The uneven distribution of $\rm cos\theta$ is due to the structure of crystals in the EMC.}\label{compare_digamma}
\end{figure*}

To select $e^{+}e^{-}\to \gamma\gamma$ events, the number of good charged tracks is required to be zero. Two neutral clusters are required to have a polar angle $|\cos\theta|$ $<$ 0.8 with the deposited energy $E_\gamma$ satisfied $0.7$ $<$ $E_\gamma$/$E_{\rm beam}$ $<$ $1.16$. The two selected photon candidates are further required to be back to back by applying the requirement $|\Delta\phi_{\gamma}|=|\phi_{\gamma1}-\phi_{\gamma2}|<2.5^{\circ}$, where $\phi_{\gamma1/2}$ are the azimuthal anlge of the photons. Figure~\ref{compare_digamma} shows the comparisons of the enegy deposition, polar angle and $\Delta\phi_{\gamma}$ distributions of two selected photons between experimental data and MC simulation at $\sqrt{s}$ = 2.2324 GeV.

The numbers of observed QED events, $N_{\rm QED}^{\rm obs}$, are obtained by event-counting after applying the event selection requirements on experimental data at different c.m. energies, individually. The detection efficiencies of signals, $\varepsilon_{\rm QED}$, are obtained by analyzing the corresponding signal MC events as done in data analysis. The cross sections of selected QED processes are calculated with the Babayaga v$3.5$ generator and the trigger efficiencies are quoted from Ref.~\cite{trigger}.

To estimate the numbers of background events, $N_{\rm QED}^{\rm bkg}$, two different methods are applied for $e^{+}e^{-}\to$ ($\gamma$)$e^{+}e^{-}$ and $e^{+}e^{-}\to \gamma\gamma$ processes, individually. In $e^{+}e^{-}\to$ ($\gamma$)$e^{+}e^{-}$ process, the numbers of background events are estimated by performing the same requirements on the background MC samples, which yields a background level of $10^{-5}$ after normalization. In $e^{+}e^{-}\to \gamma\gamma$ process, the background level is relatively large due to the hadronic process contamination. The normalized numbers of background events from $e^{+}e^{-}\to \gamma\gamma$ are estimated from the $\Delta\phi_{\gamma}$ sideband region, defined as $2.5^{\circ}$ $<$ $|\Delta\phi_{\gamma}|$ $<$ $5.0^{\circ}$. The distributions of the $\Delta\phi_{\gamma}$ sideband is supposed to be flat by analyzing the background MC samples.

Table~\ref{bkg-digam} shows input numbers used to calculate the luminosities at $\sqrt{s}$ = 2.2324 and 3.0969 GeV.

 \begin{table}[htbp]
  \centering
  \caption{Summaries of the input numbers in luminosity calculation at $\sqrt{s}$ = 2.2324 and 3.0969 GeV.}
  \begin{tabular}{c|c|c|c|c|c|c|c}
  \hline
  $\sqrt{s}$ & QED & $N_{\rm QED}^{\rm obs}$ & $N_{\rm QED}^{\rm bkg}$ & $\sigma_{\rm QED}$  & $\varepsilon_{\rm QED}$  & $\varepsilon_{\rm QED}^{\rm trig}$  &  $\mathcal{L}$  \\
   (GeV)     &  process &           &           & (nb)      &  (\%)          & (\%)                  &  (pb$^{-1}$)    \\

  \hline
  2.2324 &  ($\gamma$)$e^{+}e^{-}$    &728522  & 8  & 1476.5  & 18.74   & 100   & 2.645  \\
  2.2324 & $\gamma\gamma$   & 86974  & 1138  & 70.26  & 46.50   & 100   & 2.627  \\
  3.0969 & $\gamma\gamma$   & 36083  & 1062  & 36.59  & 46.25   & 100   & 2.069  \\
  \hline
    \end{tabular}
    \label{bkg-digam}
\end{table}

\section{\boldmath SYSTEMATIC UNCERTAINTY}
The main systematic uncertainties of the integrated luminosity are originated from the uncertainties related to the requirements on the kinematic variables, tracking efficiency, cluster reconstruction efficiency, c.m. energy, MC statistics, background estimation, trigger efficiency and generators.

For the systematic uncertainty from requirements on each kinematic variable, we re-measure the luminosity by altering the required values, $i.e.$, $|\cos\theta|$ $<$ 0.8, $|\Delta\theta_{e^{\pm}}|$ $<$ $10^{\circ}$, $|\Delta\phi_{e^{\pm}}| < 5^{\circ}$, $|\Delta\phi_{\gamma}| < 2.5^{\circ}$, $E_{e^{\pm}}/E_{\rm beam}$ $>$ 0.65 and $0.7$ $<$ $E_\gamma$/$E_{\rm beam}$ $<$ $1.16$, individually. The resultant differences of measured luminosity with respective to the nominal value are taken as the systematic uncertainty.

To study the uncertatinty of tracking efficiency, a Bhabha event sample is selected with only EMC information~\cite{3770}.
The candidate events are selected by requiring the two clusters registered in the EMC with the deposited energy larger than $0.65$ $\times$ $E_{\rm beam}$ and lied within the polar angle $|\cos\theta|$ $<$ 0.8, corresponding to the angular coverage of the barrel EMC. Since the two clusters originated from $e^{\pm}$ in the $e^{+}e^{-}$ $\to$ ($\gamma$)$e^{+}e^{-}$ candidate events are bent in the magnetic field, the two shower clusters in the $xy$-plane of the EMC are not back-to-back. $\Delta\phi_{e^{\pm}}$ is required to be in the range of $[-40^{\circ},-5^{\circ}]$ or $[5^{\circ},40^{\circ}]$ to remove the $e^{+}e^{-}$ $\to$ $\gamma\gamma$ events. We further apply the MDC information on the selected candidates, and the ratio of survived events is regarded as the tracking efficiency. The average difference on the tracing efficiency between data and signal MC simulation, 0.41\%, is taken as the systematic uncertainty.

The systematic uncertainty due to the cluster reconstruction efficiency in the EMC is determined to be $0.05\%$ for $e^{\pm}$ by comparing the cluster reconstruction efficiencies between data and signal MC (both for $e^{+}$ and $e^{-}$). Since high-energy $\gamma$ and $e^{\pm}$ behave in good approximation in the EMC, the value of $0.05\%$ is also taken as the systematic uncertainty due to the cluster reconstruction efficiency in the EMC for a single $\gamma$.

The uncertainty of c.m. energy is estimated to be 2 MeV~\cite{gaoqing}. For each energy point, an alternative MC simulation sample of 1 million events with a c.m. energy of 2 MeV above the nominal value are generated to re-estimate the detection efficiency, the results difference is regarded as the systematic uncertainty from c.m. energy.

The uncertainty of MC statistics is $0.17\%$ for the $e^{+}e^{-}$ $\to$ ($\gamma$)$e^{+}e^{-}$ process and $0.15\%$ for the $e^{+}e^{-}$ $\to$ $\gamma\gamma$ process, which is estimated by
\begin{equation}\
    \frac{1}{\sqrt{N}}\cdot{\sqrt{\frac{(1-\varepsilon)}{\varepsilon}}},
\end{equation}
where $N$ is the number of signal MC events, and $\varepsilon$ is the detection efficiency.

The rate of background events in the selected $e^{+}e^{-}$ $\to$ ($\gamma$)$e^{+}e^{-}$ candidate events is very small $(10^{-5})$. Therefore, the uncertainty due to background contamination is neglected. For $e^{+}e^{-}$ $\to$ $\gamma\gamma$ events, the rate of background events is the normalized number of selected background events in the sideband region divided by the number of signal events, which are (1.53$\pm$0.03)\% and (1.31$\pm$0.04)\% for experimental data and the MC simulation, respectively. Therefore, the difference $0.23\%$ is taken as uncertainty from background contamination.

The trigger efficiencies for barrel $e^{+}e^{-}$ $\to$ ($\gamma$)$e^{+}e^{-}$ events and $e^{+}e^{-}$ $\to$ $\gamma\gamma$ events are $100\%$ with an uncertainty of less than $0.1\%$~\cite{trigger}.

The uncertainty due to the Babayaga generator v$3.5$ is $0.5\%$ for $e^{+}e^{-}$ $\to$ ($\gamma$)$e^{+}e^{-}$, while $1.0\%$ for $e^{+}e^{-}$ $\to$ $\gamma\gamma$~\cite{babayaga}.

Systematic uncertainties at $\sqrt{s}$ = 2.2324 GeV for $e^{+}e^{-}$ $\to$ ($\gamma$)$e^{+}e^{-}$ and $e^{+}e^{-}$ $\to$ $\gamma\gamma$ are listed in Table~\ref{errors1}. Assuming all sources of systematic uncertainties are uncorrelated, the total uncertainty is calculated to be $0.7\%$ for $e^{+}e^{-}$ $\to$ ($\gamma$)$e^{+}e^{-}$ and $1.1\%$ for $e^{+}e^{-}$ $\to$ $\gamma\gamma$ by adding all the contributions in quadrature. The uncertainties related with the tracking efficiency, cluster reconstruction efficiency, trigger efficiency and generators are common between the different c.m. energy points, while others are c.m. energy dependent and are determined for the different c.m. energy points, individually.

\begin{table}[htbp]
  \centering
  \caption{Summary of systematic uncertainties at $\sqrt{s}$ = 2.2324 GeV.}
  \begin{tabular}{c|c|c}
  \hline
  \hline
  Source                                & $e^{+}e^{-}$ $\to$ ($\gamma$)$e^{+}e^{-}$    &   $e^{+}e^{-}$ $\to$ $\gamma\gamma$   \\
  \hline
  $|\cos\theta|$ $<$ 0.8                & $0.12$                             &    $0.18$                  \\
  $|\Delta\theta_{e^{\pm}}|$ $<$ $10^{\circ}$     & $0.05$                             &      -                    \\
  $|\Delta\phi_{e^{\pm}}| < 5^{\circ}$  & $0.01$                             &      -                    \\
  $|\Delta\phi_{\gamma}| < 2.5^{\circ}$ &  -                                 &    $0.07$                  \\
  $E_{e^{+}}/E_{\rm beam}$ $>$ 0.65           & $0.04$                             &      -                    \\
  $E_{e^{-}}/E_{\rm beam}$ $>$ 0.65           & $0.05$                             &      -                    \\
  $0.7$ $<$ $E_\gamma$/$E_{\rm beam}$ $<$ $1.16$          &  -                                 &    $0.10$                  \\
  Tracking efficiency                   & $0.41$                             &      -                    \\
  Cluster reconstruction                & $0.10$                             &    $0.10$                  \\
  Beam energy                           & $0.09$                             &    $0.09$                  \\
  MC statistics                         & $0.17$                             &    $0.15$                  \\
  Background estimation                 & $0.00$                             &    $0.23$                  \\
  Trigger efficiency                    & $0.10$                             &    $0.10$                  \\
  Generator                             & $0.50$                             &    $1.00$                  \\
  \hline
  Total                                 & $0.70$                             &    $1.10$                  \\
  \hline
    \end{tabular}
    \label{errors1}
\end{table}

\section{\boldmath SUMMARY}
\begin{figure}[htbp]
\begin{center}
\begin{overpic}[width=7.3cm,height=5cm,angle=0]{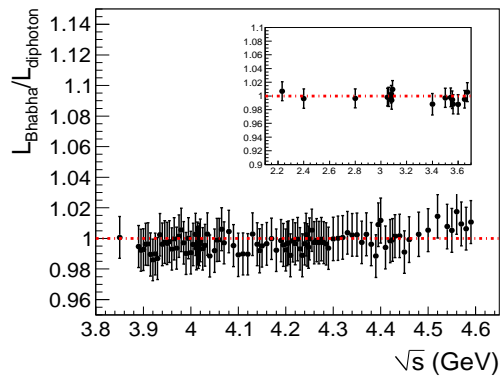}
\put(70,50){}
\end{overpic}
\end{center}
\caption{ The ratios of luminosities measured by $e^{+}e^{-}$ $\to$ ($\gamma$)$e^{+}e^{-}$ and $e^{+}e^{-}$ $\to$ $\gamma\gamma$. The major plot is for the data samples with c.m. energy larger than 3.8500 GeV, while others are shown in the insert plot, the two methods give fully compatible results within the quoted uncertainties. }\label{compare_all_ecm}
\end{figure}

By using the QED processes $e^{+}e^{-}$ $\to$ ($\gamma$)$e^{+}e^{-}$ and $e^{+}e^{-}$ $\to$ $\gamma\gamma$, the integrated luminosities have been measured for 131 data samples with c.m. energy between 2.2324 and 4.5900 GeV. The precision of integrated luminosity is around $0.7\%$ for $e^{+}e^{-}$ $\to$ ($\gamma$)$e^{+}e^{-}$, while around $1.1\%$ for $e^{+}e^{-}$ $\to$ $\gamma\gamma$. The total luminosity is 1036.3 pb$^{-1}$, and the luminosities at the individual c.m. energy point are summarized in Table~\ref{lumi-3processes}. The ratio of the measured luminosity from two process is illustrated in Fig.~\ref{compare_all_ecm}. The ratios are closed to 1 within the uncertainties, which indicates the results from the two measurements are consistent well with each other. For each energy point out of the $J/\psi$ resonance region, the luminosity measured by $e^{+}e^{-} \to$ ($\gamma$)$e^{+}e^{-}$ is more precise and thus is recommended. For energy points around $J/\psi$ (from 3.0930 to 3.1200 GeV), only the luminosities measured by $e^{+}e^{-} \to \gamma\gamma$ are obtained. The measured results are the important inputs for the physics studies, $e.g.$, R value measurement and $J/\psi$ resonance parameter measurement.

  \begin{longtable}{c|c|c}
  \caption{\label{lumi-3processes}The summaries of measured integrated luminosities from the two QED processes. The first uncertainty is statistical and the second one is systematic.}  \\
  \hline
  \endfirsthead
  \caption{(continued) The summaries of measured integrated luminosities from the two QED processes.} \\
  \hline
  \hline
  $\sqrt{s}$ (GeV)      & $e^{+}e^{-}$ $\to$ ($\gamma$)$e^{+}e^{-}$ (pb$^{-1}$) & $e^{+}e^{-}$ $\to$ $\gamma\gamma$ (pb$^{-1}$) \\
  \hline
  \endhead \hline \endfoot
  $\sqrt{s}$ (GeV)      & $e^{+}e^{-}$ $\to$ ($\gamma$)$e^{+}e^{-}$ (pb$^{-1}$) & $e^{+}e^{-}$ $\to$ $\gamma\gamma$ (pb$^{-1}$)  \\
  \hline
2.2324  &  2.645$\pm$0.006$\pm$0.020  &  2.627$\pm$0.009$\pm$0.028  \\
2.4000  & 3.415$\pm$0.007$\pm$0.024  & 3.428$\pm$0.011$\pm$0.040    \\
2.8000  & 3.753$\pm$0.008$\pm$0.026  & 3.766$\pm$0.014$\pm$0.042    \\
3.0500  & 14.893$\pm$0.030$\pm$0.103  & 14.919$\pm$0.029$\pm$0.158    \\
3.0600  & 15.040$\pm$0.030$\pm$0.131  & 15.060$\pm$0.029$\pm$0.158   \\
3.0800  & 31.019$\pm$0.060$\pm$0.189  & 30.942$\pm$0.044$\pm$0.338   \\
3.0830  & 4.740$\pm$0.011$\pm$0.029  & 4.769$\pm$0.017$\pm$0.052    \\
3.0900  & 15.709$\pm$0.031$\pm$0.099  & 15.558$\pm$0.030$\pm$0.162    \\
3.0930       &         --                           &    14.910$\pm$0.030$\pm$0.157             \\
3.0943       &         --                           &    2.143$\pm$0.011$\pm$0.023              \\
3.0952       &         --                           &    1.816$\pm$0.010$\pm$0.019              \\
3.0958       &         --                           &    2.135$\pm$0.011$\pm$0.023              \\
3.0969       &         --                           &    2.069$\pm$0.011$\pm$0.024              \\
3.0982       &         --                           &    2.203$\pm$0.011$\pm$0.023              \\
3.0990       &         --                           &    0.756$\pm$0.007$\pm$0.008              \\
3.1015       &         --                           &    1.612$\pm$0.010$\pm$0.018              \\
3.1055       &         --                           &    2.106$\pm$0.011$\pm$0.022              \\
3.1120       &         --                           &    1.720$\pm$0.010$\pm$0.019              \\
3.1200       &         --                           &    1.264$\pm$0.009$\pm$0.013              \\
3.4000  & 1.733$\pm$0.005$\pm$0.014  & 1.754$\pm$0.012$\pm$0.020    \\
3.5000  & 3.633$\pm$0.009$\pm$0.025  & 3.643$\pm$0.017$\pm$0.040    \\
3.5424  & 8.693$\pm$0.019$\pm$0.060  & 8.711$\pm$0.027$\pm$0.098    \\
3.5538  & 5.562$\pm$0.013$\pm$0.034  & 5.593$\pm$0.021$\pm$0.059    \\
3.5611  & 3.847$\pm$0.009$\pm$0.028  & 3.894$\pm$0.018$\pm$0.043    \\
3.6002  & 9.502$\pm$0.020$\pm$0.076  & 9.620$\pm$0.028$\pm$0.108    \\
3.6500  & 48.385$\pm$0.094$\pm$0.300  & 48.618$\pm$0.065$\pm$0.538    \\
3.6710  & 4.628$\pm$0.011$\pm$0.028  & 4.603$\pm$0.020$\pm$0.052   \\
3.8500  & 7.967$\pm$0.018$\pm$0.055  & 7.962$\pm$0.028$\pm$0.088    \\
3.8900  & 7.758$\pm$0.018$\pm$0.054  & 7.799$\pm$0.028$\pm$0.087    \\
3.8950  & 7.567$\pm$0.018$\pm$0.053  & 7.626$\pm$0.027$\pm$0.085    \\
3.9000  & 7.575$\pm$0.018$\pm$0.053  & 7.631$\pm$0.027$\pm$0.085    \\
3.9050  & 7.596$\pm$0.018$\pm$0.053  & 7.625$\pm$0.027$\pm$0.085    \\
3.9100  & 7.240$\pm$0.017$\pm$0.050  & 7.267$\pm$0.027$\pm$0.082    \\
3.9150  & 7.454$\pm$0.018$\pm$0.052  & 7.533$\pm$0.027$\pm$0.088    \\
3.9200  & 6.806$\pm$0.016$\pm$0.048  & 6.903$\pm$0.026$\pm$0.076   \\
3.9250  & 6.694$\pm$0.016$\pm$0.046  & 6.763$\pm$0.026$\pm$0.075    \\
3.9300  & 6.735$\pm$0.016$\pm$0.047  & 6.825$\pm$0.026$\pm$0.076    \\
3.9350  & 7.161$\pm$0.017$\pm$0.051  & 7.144$\pm$0.027$\pm$0.079    \\
3.9400  & 7.228$\pm$0.017$\pm$0.050  & 7.256$\pm$0.027$\pm$0.082    \\
3.9450  & 7.590$\pm$0.018$\pm$0.054  & 7.608$\pm$0.028$\pm$0.086    \\
3.9500  & 7.714$\pm$0.018$\pm$0.055  & 7.739$\pm$0.028$\pm$0.086   \\
3.9550  & 8.124$\pm$0.019$\pm$0.056  & 8.141$\pm$0.029$\pm$0.090    \\
3.9600  & 8.489$\pm$0.020$\pm$0.061  & 8.548$\pm$0.029$\pm$0.095   \\
3.9650  & 7.768$\pm$0.018$\pm$0.054  & 7.770$\pm$0.028$\pm$0.086    \\
3.9700  & 7.321$\pm$0.017$\pm$0.051  & 7.368$\pm$0.028$\pm$0.082    \\
3.9750  & 8.062$\pm$0.019$\pm$0.057  & 8.050$\pm$0.029$\pm$0.089    \\
3.9800  & 7.851$\pm$0.019$\pm$0.059  & 7.808$\pm$0.028$\pm$0.087    \\
3.9850  & 7.969$\pm$0.019$\pm$0.057  & 7.992$\pm$0.029$\pm$0.089    \\
3.9900  & 8.024$\pm$0.019$\pm$0.056  & 8.104$\pm$0.029$\pm$0.091    \\
3.9950  & 7.985$\pm$0.019$\pm$0.057  & 7.984$\pm$0.028$\pm$0.084   \\
4.0000  & 7.732$\pm$0.018$\pm$0.056  & 7.805$\pm$0.028$\pm$0.088   \\
4.0050  & 7.537$\pm$0.018$\pm$0.053  & 7.567$\pm$0.028$\pm$0.085    \\
4.0100  & 7.183$\pm$0.017$\pm$0.050  & 7.164$\pm$0.027$\pm$0.079   \\
4.0120  & 6.907$\pm$0.017$\pm$0.051  & 6.951$\pm$0.027$\pm$0.079   \\
4.0140  & 6.694$\pm$0.016$\pm$0.048  & 6.716$\pm$0.027$\pm$0.075    \\
4.0160  & 6.544$\pm$0.016$\pm$0.045  & 6.582$\pm$0.026$\pm$0.074   \\
4.0180  & 6.968$\pm$0.017$\pm$0.049  & 6.996$\pm$0.027$\pm$0.078    \\
4.0200  & 6.726$\pm$0.016$\pm$0.047  & 6.735$\pm$0.027$\pm$0.075   \\
4.0250  & 6.538$\pm$0.016$\pm$0.047  & 6.583$\pm$0.026$\pm$0.073    \\
4.0300  & 16.451$\pm$0.036$\pm$0.115  & 16.526$\pm$0.042$\pm$0.187    \\
4.0350  & 6.706$\pm$0.016$\pm$0.047  & 6.687$\pm$0.027$\pm$0.074   \\
4.0400  & 6.564$\pm$0.016$\pm$0.046  & 6.640$\pm$0.027$\pm$0.073   \\
4.0500  & 6.567$\pm$0.016$\pm$0.047  & 6.620$\pm$0.027$\pm$0.076   \\
4.0550  & 6.927$\pm$0.017$\pm$0.052  & 6.934$\pm$0.027$\pm$0.077    \\
4.0600  & 6.338$\pm$0.015$\pm$0.045  & 6.344$\pm$0.026$\pm$0.071    \\
4.0650  & 7.022$\pm$0.017$\pm$0.050  & 6.980$\pm$0.027$\pm$0.077   \\
4.0700  & 7.271$\pm$0.017$\pm$0.052  & 7.292$\pm$0.028$\pm$0.079    \\
4.0800  & 7.721$\pm$0.018$\pm$0.054  & 7.686$\pm$0.029$\pm$0.085    \\
4.0900  & 7.611$\pm$0.018$\pm$0.054  & 7.647$\pm$0.029$\pm$0.084    \\
4.1000  & 7.254$\pm$0.017$\pm$0.051  & 7.333$\pm$0.029$\pm$0.085    \\
4.1100  & 7.146$\pm$0.017$\pm$0.050  & 7.219$\pm$0.028$\pm$0.080    \\
4.1200  & 7.648$\pm$0.018$\pm$0.053  & 7.728$\pm$0.028$\pm$0.085    \\
4.1300  & 7.207$\pm$0.017$\pm$0.051  & 7.187$\pm$0.029$\pm$0.079    \\
4.1400  & 7.268$\pm$0.017$\pm$0.051  & 7.296$\pm$0.030$\pm$0.082    \\
4.1450  & 7.774$\pm$0.019$\pm$0.057  & 7.837$\pm$0.029$\pm$0.092   \\
4.1500  & 7.662$\pm$0.018$\pm$0.053  & 7.699$\pm$0.028$\pm$0.087    \\
4.1600  & 7.954$\pm$0.019$\pm$0.056  & 7.982$\pm$0.030$\pm$0.090   \\
4.1700  & 18.008$\pm$0.039$\pm$0.130  & 18.012$\pm$0.045$\pm$0.197    \\
4.1800  & 7.309$\pm$0.018$\pm$0.051  & 7.366$\pm$0.029$\pm$0.082   \\
4.1900  & 7.560$\pm$0.018$\pm$0.052  & 7.571$\pm$0.029$\pm$0.084    \\
4.1950  & 7.503$\pm$0.018$\pm$0.054  & 7.535$\pm$0.029$\pm$0.084    \\
4.2000  & 7.582$\pm$0.018$\pm$0.053  & 7.640$\pm$0.030$\pm$0.084    \\
4.2030  & 6.815$\pm$0.017$\pm$0.048  & 6.838$\pm$0.028$\pm$0.080    \\
4.2060  & 7.638$\pm$0.018$\pm$0.055  & 7.660$\pm$0.030$\pm$0.088    \\
4.2100  & 7.678$\pm$0.018$\pm$0.054  & 7.764$\pm$0.030$\pm$0.089   \\
4.2150  & 7.768$\pm$0.019$\pm$0.054  & 7.780$\pm$0.030$\pm$0.087    \\
4.2200  & 7.935$\pm$0.019$\pm$0.055  & 7.963$\pm$0.030$\pm$0.088    \\
4.2250  & 8.212$\pm$0.020$\pm$0.061  & 8.216$\pm$0.031$\pm$0.092   \\
4.2300  & 8.193$\pm$0.020$\pm$0.057  & 8.249$\pm$0.031$\pm$0.093    \\
4.2350  & 8.273$\pm$0.020$\pm$0.057  & 8.365$\pm$0.031$\pm$0.097   \\
4.2400  & 7.830$\pm$0.019$\pm$0.054  & 7.858$\pm$0.030$\pm$0.087    \\
4.2430  & 8.571$\pm$0.020$\pm$0.060  & 8.550$\pm$0.032$\pm$0.096   \\
4.2450  & 8.487$\pm$0.020$\pm$0.060  & 8.523$\pm$0.032$\pm$0.095    \\
4.2480  & 8.554$\pm$0.020$\pm$0.059  & 8.603$\pm$0.032$\pm$0.096    \\
4.2500  & 8.596$\pm$0.020$\pm$0.060  & 8.599$\pm$0.032$\pm$0.095    \\
4.2550  & 8.657$\pm$0.020$\pm$0.060  & 8.611$\pm$0.032$\pm$0.095    \\
4.2600  & 8.880$\pm$0.021$\pm$0.063  & 8.905$\pm$0.032$\pm$0.099    \\
4.2650  & 8.629$\pm$0.020$\pm$0.061  & 8.639$\pm$0.032$\pm$0.099    \\
4.2700  & 8.548$\pm$0.020$\pm$0.060  & 8.571$\pm$0.032$\pm$0.096   \\
4.2750  & 8.567$\pm$0.020$\pm$0.060 & 8.571$\pm$0.032$\pm$0.099   \\
4.2800  & 8.723$\pm$0.021$\pm$0.060  & 8.747$\pm$0.032$\pm$0.097    \\
4.2850  & 8.596$\pm$0.020$\pm$0.059  & 8.627$\pm$0.032$\pm$0.097   \\
4.2900  & 9.010$\pm$0.021$\pm$0.062  & 9.068$\pm$0.033$\pm$0.102    \\
4.3000  & 8.453$\pm$0.020$\pm$0.064  & 8.456$\pm$0.031$\pm$0.095    \\
4.3100  & 8.599$\pm$0.021$\pm$0.063  & 8.598$\pm$0.032$\pm$0.100   \\
4.3200  & 9.342$\pm$0.022$\pm$0.065  & 9.336$\pm$0.033$\pm$0.109    \\
4.3300  & 8.657$\pm$0.021$\pm$0.063  & 8.625$\pm$0.031$\pm$0.095    \\
4.3400  & 8.700$\pm$0.021$\pm$0.061  & 8.680$\pm$0.031$\pm$0.097    \\
4.3500  & 8.542$\pm$0.020$\pm$0.064  & 8.521$\pm$0.031$\pm$0.094   \\
4.3600  & 8.063$\pm$0.019$\pm$0.057  & 8.084$\pm$0.031$\pm$0.090    \\
4.3700  & 8.498$\pm$0.020$\pm$0.061  & 8.475$\pm$0.032$\pm$0.095   \\
4.3800  & 8.158$\pm$0.020$\pm$0.060  & 8.189$\pm$0.031$\pm$0.092    \\
4.3900  & 7.460$\pm$0.018$\pm$0.052  & 7.547$\pm$0.030$\pm$0.086    \\
4.3950  & 7.430$\pm$0.018$\pm$0.052  & 7.364$\pm$0.030$\pm$0.083    \\
4.4000  & 7.178$\pm$0.018$\pm$0.050  & 7.095$\pm$0.029$\pm$0.084    \\
4.4100  & 6.352$\pm$0.016$\pm$0.045  & 6.390$\pm$0.028$\pm$0.071    \\
4.4200  & 7.519$\pm$0.018$\pm$0.054  & 7.532$\pm$0.030$\pm$0.085   \\
4.4250  & 7.436$\pm$0.018$\pm$0.052  & 7.443$\pm$0.030$\pm$0.083   \\
4.4300  & 6.788$\pm$0.017$\pm$0.047  & 6.778$\pm$0.029$\pm$0.075    \\
4.4400  & 7.634$\pm$0.019$\pm$0.053  & 7.622$\pm$0.030$\pm$0.087   \\
4.4500  & 7.677$\pm$0.019$\pm$0.054  & 7.746$\pm$0.031$\pm$0.087    \\
4.4600  & 8.724$\pm$0.021$\pm$0.072  & 8.731$\pm$0.033$\pm$0.101    \\
4.4800  & 8.167$\pm$0.020$\pm$0.062  & 8.145$\pm$0.032$\pm$0.093    \\
4.5000  & 7.997$\pm$0.019$\pm$0.056  & 7.954$\pm$0.032$\pm$0.088    \\
4.5200  & 8.674$\pm$0.021$\pm$0.061  & 8.550$\pm$0.033$\pm$0.096    \\
4.5400  & 9.335$\pm$0.022$\pm$0.077  & 9.263$\pm$0.034$\pm$0.102    \\
4.5500  & 8.765$\pm$0.021$\pm$0.066  & 8.719$\pm$0.033$\pm$0.098    \\
4.5600  & 8.259$\pm$0.020$\pm$0.068  & 8.117$\pm$0.032$\pm$0.090    \\
4.5700  & 8.390$\pm$0.020$\pm$0.062  & 8.311$\pm$0.033$\pm$0.093   \\
4.5800  & 8.545$\pm$0.021$\pm$0.060  & 8.491$\pm$0.033$\pm$0.094    \\
4.5900  & 8.162$\pm$0.020$\pm$0.056  & 8.076$\pm$0.032$\pm$0.090    \\
  \hline
  \hline
    \end{longtable}

\acknowledgments
The BESIII collaboration thanks the staff of BEPCII and the IHEP computing center for their strong support. This work is supported in part by National Key Basic Research Program of China under Contract No. 2015CB856700; National Natural Science Foundation of China (NSFC) under Contracts Nos. 10935007, 11121092, 11125525, 11235011, 11322544, 11335008, 11375170, 11275189, 11079030, 11475164, 11475169, 11005109, 10979095, 11275211;  the Chinese Academy of Sciences (CAS) Large-Scale Scientific Facility Program; Joint Large-Scale Scientific Facility Funds of the NSFC and CAS under Contracts Nos. 11179007, U1232201, U1332201, U1532102; CAS under Contracts Nos. KJCX2-YW-N29, KJCX2-YW-N45; 100 Talents Program of CAS; INPAC and Shanghai Key Laboratory for Particle Physics and Cosmology; German Research Foundation DFG under Contract No. Collaborative Research Center CRC-1044, FOR 2359; Istituto Nazionale di Fisica Nucleare, Italy; Ministry of Development of Turkey under Contract No. DPT2006K-120470; Russian Foundation for Basic Research under Contract No. 14-07-91152; U. S. Department of Energy under Contracts Nos. DE-FG02-04ER41291, DE-FG02-05ER41374, DE-FG02-94ER40823, DESC0010118; U.S. National Science Foundation; University of Groningen (RuG) and the Helmholtzzentrum fuer Schwerionenforschung GmbH (GSI), Darmstadt; WCU Program of National Research Foundation of Korea under Contract No. R32-2008-000-10155-0.


\begin{thebibliography}{999}
   \bibitem{R_cal1}
   K.~Hagiwara {\it et al.} J. Phys. G {\bf 38}, 085003 (2011).

   \bibitem{R_cal2}
   M.~Davier {\it et al.} Eur. Phys. J. C {\bf 71}, 1515 (2011).

   \bibitem{R_cal3}
   Fred~Jegerlehner arXiv:1511.04473v2 [hep-ph] (2015).

   \bibitem{NIM:DET}
   M.~Ablikim {\it et al.} (BESIII Collaboration),
   Nucl.\ Instrum.\ Meth.\ A {\bf 614}, 345 (2010).

   \bibitem{babayaga}
   G.~Balossini {\it et al.} Nucl. Phys. B {\bf 758}, 227 (2006).

   \bibitem{KKMC}
   S.~Jadach, B.~F.~L.~Ward and Z.~Was, Comp.\ Phys.\ Commun.\  {\bf 130}, 260 (2000); Phys.\ Rev.\ D {\bf 63}, 113009 (2001).

   \bibitem{LUND2}
   B.~Andersson, The Lund Model, Cambridge University Press, 1998.


   \bibitem{bestwogam}
   S.~Nova, A.~Olchevski and T.~Todorov (DELPHI collaboration), DELPHI 90-35 PROG {\bf 152} (1990).

   \bibitem{trigger}
   N.Berger, K.~Zhu et al. Chin.\ Phys.\ C {\bf 34}, 1779 (2010).

   \bibitem{3770}
   M.~Ablikim {\it et al.} (BESIII Collaboration),
   Chin.\ Phys.\ C {\bf 37}, 123001 (2013).

   \bibitem{gaoqing}
   M.~Ablikim {\it et al.} (BESIII Collaboration),
   Chin.\ Phys.\ C {\bf 39}, 093001 (2015).
\end{thebibliography}
\end{document}